# Real-time kinematic positioning of LEO satellites using a single-frequency GPS receiver

Pei Chen, Jian Zhang, and Xiucong Sun (✉)

*School of Astronautics, Beihang University, Beijing 100191, China*

*Tel.:+86-10-82316535, e-mail: sunxiucong@gmail.com*

**Abstract** Due to their low cost and low power consumption, single-frequency GPS receivers are considered suitable for low-cost space applications such as small satellite missions. Recently, requirements have emerged for real-time accurate orbit determination at sub-meter level in order to carry out onboard geocoding of high-resolution imagery, open-loop operation of altimeters and radio occultation. This study proposes an improved real-time kinematic positioning method for LEO satellites using single-frequency receivers. The C/A code and L1 phase are combined to eliminate ionospheric effects. The epoch-differenced carrier phase measurements are utilized to acquire receiver position changes which are further used to smooth the absolute positions. A kinematic Kalman filter is developed to implement kinematic orbit determination. Actual flight data from China's small satellite SJ-9A are used to test the navigation performance. Results show that the proposed method outperforms traditional kinematic positioning method in terms of accuracy. A 3D position accuracy of 0.72 m and 0.79 m has been achieved using the predicted portion of IGS ultra-rapid products and broadcast ephemerides, respectively.

## Introduction

The high accuracy, global coverage and three-dimensional nature of the Global Positioning System (GPS) has led to a wide use of GPS receivers not only for terrestrial applications but also for space applications, such as orbit determination of low-earth-orbiting (LEO) satellites. The GPS measurements support both ground-based high-precision orbit determination and onboard real-time navigation tasks (Zehentner and Mayer-Gürr 2016; van den Ijssel et al. 2015; Montenbruck et al. 2013). In general, centimeter-level accuracy is achieved using ionospheric-free carrier phase observations from dual-frequency receivers. With the improvement of measurement accuracy, single-frequency receivers have been proposed in recent years for meter or sub-meter level orbit determination (Wang et al. 2015; Montenbruck et al. 2012; Peng and Wu 2012; Bock et al. 2009; Hwang and Born 2005). The use of single-frequency receivers can save costs and reduce energy consumption. They are preferable for small satellites or space missions requiring lower orbital accuracy.

The ionospheric propagation delay is considered a major obstacle to single-frequency precise orbit determination (POD). The GRAPHIC (GRoup And Phase Ionospheric Correction) method can be used to eliminate the first-order ionospheric effects (Yunck 1993), and sub-meter level position accuracies have been demonstrated for several satellites (Hwang et al. 2011; Bock et al. 2009). The GRAPHIC method uses the fact that the C/A code and L1 phase experience the same ionospheric delay but with opposite sign. Unlike the ionospheric-free P1/P2 pseudorange combination, the GRAPHIC observation represents a heavily biased one-way range which cannot be used for single point positioning (Montenbruck 2003). Actually, the GRAPHIC observation is similar to the ionospheric-free L1/L2 carrier phase combination but with larger observation noise due to the incorporation of C/A code measurements.

This study investigates the use of the GRAPHIC processing methodology for GPS single-frequency orbit determination of LEO satellites for real-time applications. The motivation originates from an emerging requirement for real-time precise orbits with sub-meter accuracy in spacecraft operations such as onboard geocoding of high-resolution imagery (Wermuth et al. 2012), open-loop operation of altimeters (Jayles et al. 2010) and radio occultation (Montenbruck et al. 2013). The single-frequency receivers are compatible with onboard needs of these missions, provided that precise GPS ephemerides are available (Montenbruck et al. 2012). A GRAPHIC-based dynamical filter for real-time single-frequency orbit determination was developed in Montenbruck and Ramos-Bosch (2008). GPS measurements collected from various LEO satellites including CHAMP, GRACE, TerraSAR-X, ICEsat, SAC-C, and MetOp were used to test the algorithm. Position accuracies ranging from 0.48 m to 0.69 m (3D rms) were achieved using broadcast ephemerides, and decimeter level accuracies were obtained using JPL real-time ephemerides. The GRAPHIC filter algorithm was further adapted to be implemented on the PROBA-2 microsatellite which carried a Phoenix-XNS single-frequency receiver for real-time navigation. An in-flight accuracy of 1.1 m (3D rms) was achieved using broadcast ephemerides, and the potential for accuracy improvement to 0.7 m was also demonstrated on ground with a refined filter tuning.

The sub-meter level accuracies achieved for real-time single-frequency orbit determination mentioned above are a result of using dynamical filtering technique which reduces the measurement noises and improves robustness (Goldstein et al 2001; Kroes 2006). In contrast to GPS dynamic orbit determination, the kinematic method estimates the satellite's position coordinates epoch-by-epoch and requires no a priori knowledge of the spacecraft motion (Weinbach and Schön 2013; Montenbruck 2003; Bisnath and Langley 2002). The kinematic approach can be applied to a wide range of situations and is of particular interest for maneuvering spacecraft and reentry vehicles due to its purely geometrical nature. Moreover, the computational complexity is significantly reduced compared to the dynamical filtering technique. However, the kinematic approach is sensitive to measurement outliers and bad viewing geometry. Thus its practical use is restricted.

The present study proposes an improved kinematic orbit determination method which utilizes epoch-differenced carrier phase measurements to smooth GRAPHIC-derived positions for real-time single-frequency orbit determination of LEO satellites. The epoch-differencing of the carrier phase eliminates the ambiguity term as well as the major part of ionospheric effects and can be used to obtain highly accurate displacement information between two adjacent epochs. The displacement will constrain the kinematic positions estimated from GRAPHIC observations, similar to the orbital motion information in dynamic approach. In addition, code outliers and carrier phase cycle slips can be detected using epoch-differenced observations. Thus the robustness is enhanced compared to the traditional kinematic method.

In the following chapter, the mathematical models of GPS observations and linear combinations for single-frequency orbit determination are described. The real-time GPS orbit and clock products are introduced next, and the observation errors due to inaccuracies of the real-time products are analyzed. Afterwards, the detailed algorithm of the improved kinematic orbit determination method is presented. The proposed method is finally tested with actual flight data from the single-frequency GPS receiver onboard the China's SJ-9A (ShiJian-9A) low-cost small satellite (Zhao et al. 2013), where both the broadcast ephemerides and the predicted portion of the IGS ultra-rapid ephemerides are tested.

**GPS single-frequency observation model**

In the case of single-frequency GPS receivers, only the C/A code and L1 carrier phase are available. The simplified observation equations (in meters) are given as follows (Leick et al. 2015)

$$C(t) = \rho_r^s(t) + c\left(\delta t_r(t) - \delta t^s(t - \tau_r^s)\right) + I(t) + \varepsilon_C(t) \tag{1}$$

$$L(t) = \rho_r^s(t) + c\left(\delta t_r(t) - \delta t^s(t - \tau_r^s)\right) - I(t) + A_L + \varepsilon_L(t) \tag{2}$$

where $C(t)$ and $L(t)$ denote the C/A code and L1 phase at epoch $t$, $\rho_r^s = \|\mathbf{x}_r(t) - \mathbf{x}^s(t - \tau_r^s)\|$ the receiver-satellite geometric range, $\mathbf{x}_r(t)$ and $\delta t_r(t)$ are receiver antenna phase center position and clock offset at the signal reception time, $\mathbf{x}^s(t - \tau_r^s)$ and $\delta t^s(t - \tau_r^s)$ are satellite antenna phase center position and clock offset at the signal transmitting time, $\tau_r^s$ is the signal travelling time, $c$ is the vacuum speed of light, $I(t)$ is the ionospheric propagation delay, $A_L$ is the non-integer ambiguity term (unit m), and $\varepsilon_C(t)$ and $\varepsilon_L(t)$ are the random thermal noises. The relativistic effects, the phase wind-up, as well as the hardware delays are not discussed here. The treatment of these errors could be found in Kouba (2009) and Sterle et al (2015).

The GRAPHIC combination eliminates the dominant error source in single-frequency GPS measurements, i.e., ionospheric propagation delay, and is defined as

$$\begin{aligned} G(t) &\triangleq \frac{1}{2}\left(C(t) + L(t)\right) \\ &= \rho_r^s(t) + c\left(\delta t_r(t) - \delta t^s(t - \tau_r^s)\right) + A + \varepsilon_G(t) \end{aligned} \tag{3}$$

where $A$ represents $A_L/2$ and $\varepsilon_G(t)$ represents the GRAPHIC observation noise. The standard deviation of $\varepsilon_G(t)$ is

$$\sigma_G = \frac{1}{2}\sqrt{\sigma_C^2 + \sigma_L^2} \approx \frac{1}{2}\sigma_C \tag{4}$$

where $\sigma_C$ and $\sigma_L$ are standard deviations of code and phase noises. The code noise is reduced by a factor of 2 in the GRAPHIC observation. However, a bias term which equals half of the carrier phase ambiguity is involved and must be solved along with receiver position and clock offset. It should be noted that there is a linear dependency between the receiver clock offset and the GRAPHIC ambiguity. That is, only the change of clock offset values with respect to the reference epoch can be estimated. To address this singularity problem, a priori value of clock offset at the reference epoch will be used and fixed in the estimation process.

For continuous GPS tracking, epoch-differenced observations can be formed between two observations of the same type at two adjacent epochs as follows

$$\begin{aligned}\nabla_{t_1 t_2} C &\triangleq C(t_2) - C(t_1) \\ &= \nabla_{t_1 t_2}\rho_r^s + c\left(\nabla_{t_1 t_2}\delta t_r - \nabla_{t_1 t_2}\delta t^s\right) + \nabla_{t_1 t_2} I + \nabla_{t_1 t_2}\varepsilon_C\end{aligned} \tag{5}$$

$$\begin{aligned}\nabla_{t_1 t_2} L &\triangleq L(t_2) - L(t_1) \\ &= \nabla_{t_1 t_2}\rho_r^s + c\left(\nabla_{t_1 t_2}\delta t_r - \nabla_{t_1 t_2}\delta t^s\right) - \nabla_{t_1 t_2} I + \nabla_{t_1 t_2}\varepsilon_L\end{aligned} \tag{6}$$

where the epoch-differencing operator is defined as $\nabla_{t_1 t_2}(*) = (*)(t_2) - (*)(t_1)$. The ambiguity term in (2) cancel out if no cycle slips occur. The ionospheric errors are highly correlated and the largest part of ionospheric path delay can also be eliminated. In the case of a short epoch-differencing interval of 1 or 2 seconds, which corresponds to a separation distance of about 10 kilometers for LEO satellites, the residual term $\nabla_{t_1 t_2} I$ is on the order of centimeters (Chen et al 2015; Kroes 2006). When the epoch-differencing interval increases to 5 or 10 seconds, the ionospheric path delay residual could reach decimeter level. This will be discussed in the results section.

Let $\sigma_{\nabla I}$ denote the standard deviation of $\nabla_{t_1 t_2} I$. Define $\varepsilon_{\nabla C} = \nabla_{t_1 t_2} I + \nabla_{t_1 t_2}\varepsilon_C$ and $\varepsilon_{\nabla L} = -\nabla_{t_1 t_2} I + \nabla_{t_1 t_2}\varepsilon_L$. The standard deviations of $\varepsilon_{\nabla C}$ and $\varepsilon_{\nabla L}$ respectively are

$$\sigma_{\nabla C} = \sqrt{\sigma_{\nabla I}^2 + 2\sigma_C^2} \tag{7}$$

$$\sigma_{\nabla L} = \sqrt{\sigma_{\nabla I}^2 + 2\sigma_L^2} \qquad (8)$$

The noise level of $\nabla_{t_1 t_2} L$ is much lower than that of $\nabla_{t_1 t_2} C$. Large code outliers or significant carrier phase cycle slips can be easily detected from the following geometry-free difference between epoch-differenced code and phase observations

$$\nabla_{t_1 t_2} C - \nabla_{t_1 t_2} L = \varepsilon_{\nabla C} - \varepsilon_{\nabla L} \qquad (9)$$

where only the noise terms remain after the differencing operation. The criterion for the detection is given as follows

$$\left| \nabla_{t_1 t_2} C - \nabla_{t_1 t_2} L \right| > \sqrt{3} \sqrt{\sigma_{\nabla C}^2 + \sigma_{\nabla L}^2} \qquad (10)$$

with a tail probability of 0.3%.

In the orbit determination section, the GRAPHIC model (3) will be used to estimate the receiver's absolute position and clock offset, whereas the epoch-differenced carrier phase model (6) will be used to estimate the variations of receiver position and clock offset between two adjacent epochs.

**Real-time orbit and clock products**

With ionospheric effects eliminated through the GRAPHIC combination, the GPS orbit and clock information becomes the key factor influencing the accuracy of real-time single-frequency navigation. In general, there are two sets of GPS ephemerides available for orbit determination, the broadcast ephemerides and the precise ephemerides. The broadcast ephemerides are generated and controlled by the GPS operational control segment (OCS) and transmitted in the navigation message (van Dierendonck et al 1978). The precise ephemerides are provided by the International GNSS Service (IGS) and its analysis centers, including the final, the rapid, and the ultra-rapid products (Dow et al. 2009). In order to satisfy real-time users with more precise products, the IGS now provides real-time service (RTS) products, which are streamed over the Internet using Networked Transport of RTCM via Internet Protocol (Hadas and Bosy 2015; Elsobeiey and Al-Harbi 2016).

Among these orbit and clock products, only the broadcast ephemerides, the predicted portion of the IGS ultra-rapid products, and the IGS RTS are available for real-time users. The RTS products are currently only streamed over the Internet. Continuous uploading of RTS products to an orbiting LEO satellite requires a network of globally distributed ground stations and brings additional burdens to satellite communication and mission operation. In contrast, the ultra-rapid products are updated four times per day, at 03:00, 09:00, 15:00, and 21:00 UTC. The observed orbits have an initial latency of 3 hours at the release time. From release to uploading, an additional time ranging from a few minutes to one orbit period (~1.5 hours) will be required, which depends on the relative position between the LEO user satellite and the ground station at the release time. Thus a scheduled uploading with an average interval of 6 hours would be sufficient for real-time orbit determination. Within this study, only the broadcast ephemerides and the predicted portion of the IGS ultra-rapid products are used for real-time navigation.

The inaccuracy of real-time ephemerides will introduce navigation errors. The Signal-In-Space Range Error (SISRE) describes a global root-mean-square (rms) value of range errors due to GPS orbit and clock errors (Warren and Raquet 2003; Montenbruck et al 2005). In order to investigate the time-varying characteristics of GPS orbit and clock errors, this study introduces the concept of Equivalent Range Error (ERE) which describes location dependent, instantaneous projected errors. The ERE is computed by

$$\text{ERE}(t) = \left\| \boldsymbol{x}_r(t) - \boldsymbol{x}^s(t - \tau_r^s) \right\| - \left\| \boldsymbol{x}_r(t) - \boldsymbol{x}_{RT}^s(t - \tau_r^s) \right\| \\ + c\left( \delta t_{RT}^s(t - \tau_r^s) - \delta t^s(t - \tau_r^s) \right) \quad (11)$$

where $\boldsymbol{x}_{RT}^s(t - \tau_r^s)$ and $\delta t_{RT}^s(t - \tau_r^s)$ are satellite antenna phase center and clock correction provided by real-time ephemerides, $\boldsymbol{x}^s(t - \tau_r^s)$ and $\delta t^s(t - \tau_r^s)$ are true values. It is noted that the IGS products refer to satellite's center of mass and need to be corrected with absolute phase center offsets. Define $\Delta \boldsymbol{x}^s = \boldsymbol{x}_{RT}^s - \boldsymbol{x}^s$ and $\Delta \delta t^s = \delta t_{RT}^s - \delta t^s$. A Taylor series expansion of (11) truncated at the first-order term is given as follows

$$\text{ERE}(t) \approx -\boldsymbol{e}_r^s \cdot \Delta \boldsymbol{x}_{RT}^s + c\Delta \delta t^s \quad (12)$$

where $\boldsymbol{e}_r^s$ is the line-of-sight vector from the receiver to the GPS satellite.

The time-varying characteristics of EREs for the real-time orbit and clock products are illustrated in Figure 1. A subset of GPS data from the SJ-9A satellite have been processed and the GPS satellite PRN 12 (Block IIR-M) is investigated. The IGS final products are used as reference. The absolute phase center offsets have been corrected for IGS orbits. In order to account for time scale biases for different ephemeris products, an ensemble clock difference is computed at each epoch from the average clock differences of all the satellites, and the individual clock offset differences are corrected for this ensemble average (Montenbruck et al. 2015). The broadcast ephemerides contain large orbit and clock correction errors. The corresponding ERE curve (blue solid line) can be decomposed by a drifting bias and random noise. However, the predicted portion of IGS ultra-rapid products are mainly governed by clock correction errors. The corresponding ERE curve (red dashed line) resembles a constant bias contaminated by random noise. Similar phenomena have been found for the other GPS satellites, although not presented here. The constant part of ERE will be absorbed by the ambiguity term and the receiver clock offset. In order to incorporate the ERE's random noise part, the GRAPHIC ambiguity is modeled as the following stochastic process

$$A(t_k) = A(t_{k-1}) + v_k \qquad (13)$$

where $v_k$ is assumed to be zero-mean Gaussian noise, and its standard deviation can be pre-determined from statistical analysis.

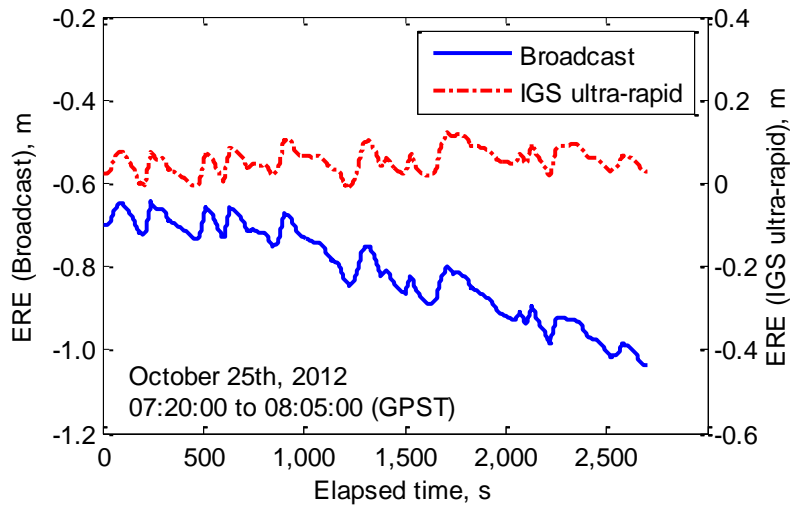

**Fig. 1** Time-varying EREs of broadcast ephemerides (blue solid) and the predicted portion of IGS ultra-rapid products (red dashed) of PRN 12. The IGS final products are used as reference

The effects of orbit and clock correction errors on epoch-differenced observations can be evaluated by the epoch-differenced ERE

$$\nabla_{t_1 t_2} \text{ERE} = \text{ERE}(t_2) - \text{ERE}(t_1)$$
$$\approx -\left(\left(\nabla_{t_1 t_2} \boldsymbol{e}_r^s\right) \cdot \Delta \boldsymbol{x}_{RT}^s(t_1) + \boldsymbol{e}_r^s(t_1) \cdot \left(\nabla_{t_1 t_2} \Delta \boldsymbol{x}_{RT}^s\right)\right) + c\left(\nabla_{t_1 t_2} \Delta \delta t^s\right) \quad (14)$$

where $\nabla_{t_1 t_2} \boldsymbol{e}_r^s$ is the variation of the line-of-sight vector. In the case of short time interval of 1 or 2 seconds, $\left\|\nabla_{t_1 t_2} \boldsymbol{e}_r^s\right\| \approx \frac{10 \text{ km}}{26000 \text{ km}} = 3.8 \times 10^{-4}$. The time differences of $\Delta \boldsymbol{x}_{RT}^s$ and $c\Delta \delta t^s$ are usually quite small due to the smoothness of the predicted GPS orbits and clocks, which are on the orders of millimeters. Thus the standard deviation of $\nabla_{t_1 t_2} \text{ERE}$ can be estimated to be on the order of millimeters. This has been validated by the actual values of time differences of the ERE data which were used for plotting Figure 1. The rms values of $\nabla_{t_1 t_2} \text{ERE}$ for the broadcast ephemerides and the IGS ultra-rapid (predicted) product are 0.934 mm and 0.923 mm, respectively. The rms of $\nabla_{t_1 t_2} \text{ERE}$ can be used as the standard deviation of the ambiguity noise. The magnitude of $\nabla_{t_1 t_2} \text{ERE}$ is much smaller than the epoch-differenced carrier phase noise level and will be neglected in the estimation of variations of receiver position and clock offset.

**Improved kinematic real-time navigation**

In traditional kinematic navigation, the epoch-wise receiver positions, the epoch-wise receiver clock offsets, as well as the constant ambiguity terms which are associated with continuous, uninterrupted arcs of carrier phase tracking are estimated by a recursive least-squares filter in a sequential manner. Without dynamical constraints or data preprocessing, the estimated positions will be sensitive to observation noise, outliers and cycle slips, especially for single-frequency navigation where code measurements are used.

In order to reduce the observation noise and enhance the robustness at the presence of code outliers or carrier phase cycle slips, we introduce epoch-differenced observations for kinematic navigation and develop a kinematic Kalman filter to implement the state estimation. Figure 2 illustrates the procedure of the proposed filter for real-time single-frequency kinematic orbit determination, where the "+" superscript denotes a posteriori estimation and the "−" superscript denotes a priori estimation. The filter in one cycle consists of a time-update stage and a measurement-update stage. At the time-update stage, a check of visible GPS satellite change is first carried out to address the cases where new satellites are available and/or old satellites disappear. Next, the epoch-differenced observations are processed to detect outliers and cycle slips, see (10). After removal of erroneous measurements, the epoch-differenced carrier phases are then used to estimate the variations of receiver position and clock offset from previous epoch to current epoch, and the ambiguities are propagated according to (13). At the measurement-update stage, the observed-minus-computed GRAPHIC measurements are used to estimate the absolute receiver position, clock offset, and ambiguities.

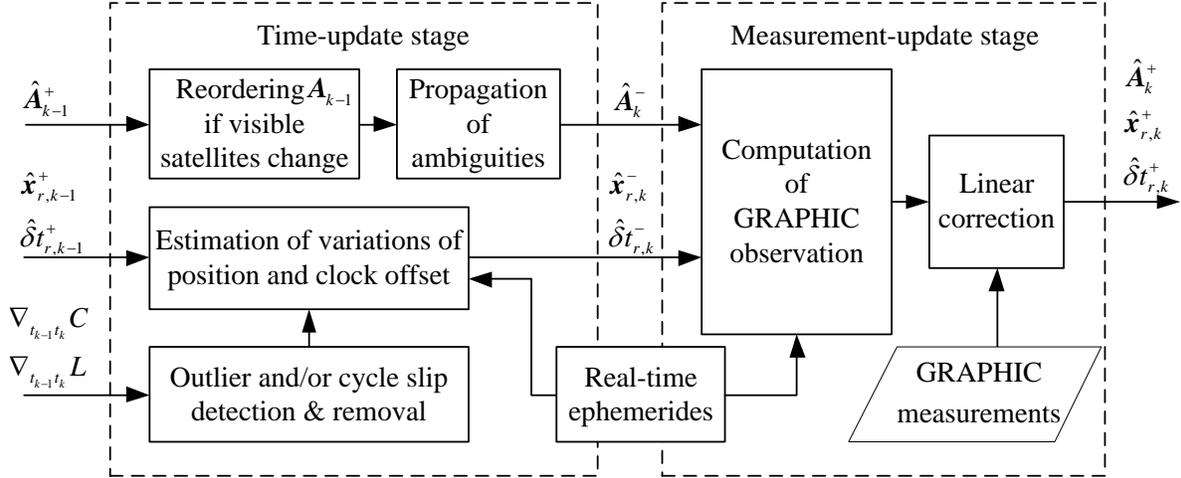

**Fig. 2** Block diagram of one cycle in the kinematic Kalman filter

Estimation of variations of receiver position and clock offset

The epoch-differenced carrier phase measurements contain the time-varying information of the rigorous "pseudorange", which refers to topocentric satellite distance plus clock offset terms, from which the variations of receiver position and clock offset can be estimated with at least four continuous observed GPS satellites in the absence of cycle slips. Consider a LEO satellite continuously tracks $n$ ($n \geq 4$) GPS satellites from epoch $t_{k-1}$ to $t_k$. The epoch-differenced carrier phase observations corrected with GPS satellite clock corrections can be cast into the following vectorial model

$$\boldsymbol{y}_k \triangleq \begin{bmatrix} \nabla_{t_{k-1}t_k} L_r^1 + c\nabla_{t_{k-1}t_k} \delta t_{RT}^1 \\ \nabla_{t_{k-1}t_k} L_r^2 + c\nabla_{t_{k-1}t_k} \delta t_{RT}^2 \\ \vdots \\ \nabla_{t_{k-1}t_k} L_r^n + c\nabla_{t_{k-1}t_k} \delta t_{RT}^n \end{bmatrix}$$

$$= \begin{bmatrix} \left\| \hat{\boldsymbol{x}}_{r,k-1}^+ + \nabla_{t_{k-1}t_k}\boldsymbol{x}_r - \boldsymbol{x}_{RT,k}^1 \right\| - \left\| \hat{\boldsymbol{x}}_{r,k-1}^+ - \boldsymbol{x}_{RT,k-1}^1 \right\| \\ \left\| \hat{\boldsymbol{x}}_{r,k-1}^+ + \nabla_{t_{k-1}t_k}\boldsymbol{x}_r - \boldsymbol{x}_{RT,k}^2 \right\| - \left\| \hat{\boldsymbol{x}}_{r,k-1}^+ - \boldsymbol{x}_{RT,k-1}^2 \right\| \\ \vdots \\ \left\| \hat{\boldsymbol{x}}_{r,k-1}^+ + \nabla_{t_{k-1}t_k}\boldsymbol{x}_r - \boldsymbol{x}_{RT,k}^n \right\| - \left\| \hat{\boldsymbol{x}}_{r,k-1}^+ - \boldsymbol{x}_{RT,k-1}^n \right\| \end{bmatrix} + \begin{bmatrix} c \\ c \\ \vdots \\ c \end{bmatrix} \nabla_{t_{k-1}t_k} \delta t_r + \boldsymbol{\varepsilon}_{\nabla L} \quad (15)$$

where $\hat{\boldsymbol{x}}_{r,k-1}^+$ denote the a posteriori estimated receiver position at $t_{k-1}$ referring to signal reception time, $\boldsymbol{x}_{RT,k-1}^i$ and $\boldsymbol{x}_{RT,k}^i$ $(i=1,2,...,n)$ denote the $i$th GPS satellite's positions at $t_{k-1}$ and $t_k$ referring to signal transmitting time and are provided in the real-time ephemerides, and $\delta t_{RT}^i$ $(i=1,2,...,n)$ is the real-time clock correction of the $i$th GPS satellite. $\boldsymbol{\varepsilon}_L$ is the observation noise vector and its covariance matrix is denoted by $\boldsymbol{R}_{\nabla L,k} = \boldsymbol{I}_{n\times n}\sigma_{\nabla L}$, where $\boldsymbol{I}_{n\times n}$ is an identity matrix.

The unknowns in (15) include three kinematic receiver position changes $\nabla_{t_{k-1}t_k}\boldsymbol{x}_r$ and one receiver clock offset change $\nabla_{t_{k-1}t_k}\delta t_r$. The estimation error of $\hat{\boldsymbol{x}}_{r,k-1}^+$, denoted as $\tilde{\boldsymbol{x}}_{r,k-1}$, will affect the equation by the relation $\left(\nabla_{t_{k-1}t_k}\boldsymbol{e}_r^s\right)\cdot\tilde{\boldsymbol{x}}_{r,k-1}$, which is easily obtained by linearization of (15) around $\hat{\boldsymbol{x}}_{r,k-1}$. This error term is many times smaller than the observation noise. As mentioned earlier, the epoch-differenced ERE is on the order of millimeters. Thus the inaccuracies of real-time ephemerides can also be neglected.

The nonlinear equation (15) can be solved using an iterated least-squares estimator. The iteration will involve the partial derivatives of $\boldsymbol{y}$ with respect to the unknowns, which are given as follows

$$\boldsymbol{H}_{\nabla,k} = \begin{bmatrix} -\left(\boldsymbol{e}_{r,k}^1\right)^T, & c \\ -\left(\boldsymbol{e}_{r,k}^2\right)^T, & c \\ \vdots & \vdots \\ -\left(\boldsymbol{e}_{r,k}^n\right)^T, & c \end{bmatrix}_{n\times 4} \tag{16}$$

The covariance matrix of the estimates is

$$\boldsymbol{P}_{\nabla,k} = \left(\boldsymbol{H}_{\nabla,k}^T \boldsymbol{R}_{\nabla L,k}^{-1} \boldsymbol{H}_{\nabla,k}\right)^{-1} \tag{17}$$

$\boldsymbol{P}_{\nabla,k}$ will be used for the measurement update of receiver position and clock offset.

Absolute position, clock offset and ambiguity estimation

A priori estimates of the receiver position, clock offset and ambiguities at epoch $t_k$ can be obtained from the following state equations,

$$\hat{\boldsymbol{x}}_{r,k}^- = \hat{\boldsymbol{x}}_{r,k-1}^+ + \nabla_{t_{k-1}t_k}\hat{\boldsymbol{x}}_r \tag{18}$$

$$\delta\hat{t}_{r,k}^- = \delta\hat{t}_{r,k-1}^+ + \nabla_{t_{k-1}t_k}\delta\hat{t}_r \tag{19}$$

$$\hat{\boldsymbol{A}}_k^- = \hat{\boldsymbol{A}}_{k-1}^+ \tag{20}$$

where $\nabla_{t_{k-1}t_k}\hat{\boldsymbol{x}}_r$ and $\nabla_{t_{k-1}t_k}\delta\hat{t}_r$ are the estimated variations of position and clock offset from $t_{k-1}$ to $t_k$. The associated covariance of these a priori estimates are given by

$$\boldsymbol{P}_k^- = \boldsymbol{P}_{k-1}^+ + \begin{bmatrix} \boldsymbol{P}_{\nabla,k} & \boldsymbol{0}_{4\times n} \\ \boldsymbol{0}_{n\times 4} & \boldsymbol{Q}_A \end{bmatrix} \tag{21}$$

where $\boldsymbol{Q}_A$ is the covariance of the ambiguity random-walk noise. In the case of newly observed satellites, the carrier-minus-code observations can be used to provide a priori estimates for the newly added ambiguities.

At the measurement-update stage, the GRAPHIC observations are first linearized around at these a priori estimates

$$z_k \triangleq \begin{bmatrix} G_r^1 - \|\hat{x}_{r,k}^- - x_{RT,k}^1\| - c\delta\hat{t}_{r,k}^- + c\delta t_{RT}^1 \\ G_r^2 - \|\hat{x}_{r,k}^- - x_{RT,k}^2\| - c\delta\hat{t}_{r,k}^- + c\delta t_{RT}^2 \\ \vdots \\ G_r^n - \|\hat{x}_{r,k}^- - x_{RT,k}^n\| - c\delta\hat{t}_{r,k}^- + c\delta t_{RT}^n \end{bmatrix} - \hat{A}_k^-$$
$$= H_{\nabla,k}\begin{bmatrix}\Delta x_{r,k} \\ \Delta \delta t_{r,k}\end{bmatrix} + \Delta A + \varepsilon_G \qquad (22)$$

where $z_k$ comprises the observed-minus-computed GRAPHIC observations, $\Delta x_{r,k}$, $\Delta \delta t_{r,k}$ and $\Delta A$ represent small errors, and $\varepsilon_G$ represent the GRAPHIC noise vector and its covariance matrix is denoted by $R_{G,k} = I_{n \times n}\sigma_G$. Define $H_k = [H_{\nabla,k}, \; I_{n \times n}]$. The estimates of the receiver position, clock offset as well as ambiguities and the associated covariance can be updated as follows

$$\begin{bmatrix}\hat{x}_{r,k}^+ \\ \delta\hat{t}_{r,k}^+ \\ \hat{A}_k^+\end{bmatrix} = \begin{bmatrix}\hat{x}_{r,k}^- \\ \delta\hat{t}_{r,k}^- \\ \hat{A}_k^-\end{bmatrix} + K_k z_k \qquad (23)$$

$$P_k^+ = (I - K_k H_k)P_k^-(I - K_k H_k)^T + K_k R_{G,k} K_k^T \qquad (24)$$

where $K_k$ is the filter gain and is given by

$$K_k = P_k^- H_k^T (H_k P_k^- H_k^T + R_{G,k})^{-1} \qquad (25)$$

So far, all of the unknown parameters have been estimated sequentially.

The kinematic Kalman filter offers a better way for real-time single-frequency navigation than traditional kinematic method through using epoch-differenced carrier phases to provide pseudo "dynamic" information. It should be pointed out that the kinematic Kalman filter cannot output velocity estimation and cannot deal with data outages, since no real-world dynamic models are used. A summary of similarities and differences between the proposed and traditional techniques for single-frequency kinematic orbit determination is provided in Table 1.

**Table 1** Comparison between the proposed and traditional kinematic single-frequency orbit determination methods

| Item | Traditional method | Proposed method |
| --- | --- | --- |
| GPS measurement model | Ionospheric-free L1 code-carrier combination (GRAPHIC) | Ionospheric-free L1 code-carrier combination (GRAPHIC) |
|  |  | Epoch-differenced carrier phase |
|  | Broadcast ephemerides | Broadcast ephemerides/IGS ultra-rapid (predicted) products |
|  | Constant ambiguity | Stochastic ambiguity |
| Orbital dynamic model | Not incorporated | Not incorporated |
| Ionospheric effects | Not considered | Not considered for GRAPHIC |
|  |  | Considered for epoch-differenced carrier phase |
| Reference frame | WGS84 | WGS84/ITRF |
| Estimation technique | Recursive least-squares filter | Kinematic Kalman filter |
| Information for outlier/cycle slip detection | Measurement residual | Geometry-free difference between epoch-differenced code and phase |

**SJ-9A flight data processing results**

The kinematic Kalman filter described above is implemented in an orbit determination software (MATLAB language) developed by the Spacecraft Simulation Technology (SST) laboratory at Beihang University. The real GPS data collected from the SJ-9A satellite have been used to study the navigation performance in an offline environment.

Mission and data set description

The SJ-9 mission objective is to test China's new space technologies such as dual-spacecraft formation flying, high-performance small camera, and electronic propulsion system. The SJ-9A/B satellites were launched on October 14, 2012 into a sun synchronous orbit. SJ-9A is based on the mature CAST-2000 platform and the satellite mass is about 790 kg. SJ-9B is based on the new CAST-100 platform and the mass is about 260 kg. The pictures of the two satellites are presented in Zhao et al. (2013). Each satellite is equipped with two 12-channel single-frequency GPS receivers for absolute real-time navigation and operation control. The receivers were manufactured by the China Academy of Space Technology (CAST) and provide C/A code and L1 carrier phase measurements. The receivers on the same satellite are connected to a common antenna to perform zero-baseline test. In addition, carrier phase differential GPS measurements from the two satellites are used to provide high accuracy inter-satellite baseline solutions (Chen et al 2015). However, the absolute navigation performance is the only concern in this study.

A 7-hour GPS data set from the main receiver onboard the SJ-9A satellite covering the time from 06:00:00 (GPS time) to 13:00:00 (GPS time) on October 25, 2012 has been used for the test. The GPS data from SJ-9B are not used because its antenna position is not provided and precise reference orbits are difficult to obtain. The orbital height of SJ-9A on this day was about 650 km. This indicates that the satellite experiences relatively smaller ionospheric path delays than those at lower orbits, such as CHAMP and GRACE. The raw GPS measurements are stored in a binary file at a sampling interval of 1 s. Zero-baseline test has shown a code noise of 60 cm and a carrier phase noise of 1 mm. This implies a noise level of 30 cm for GRAPHIC observations. It is noted that multipath or group delay variations which cannot be detected from zero-baseline test may degrade the accuracy of measurements.

In addition to the mandatory GPS measurements, the spacecraft attitude data as well as the antenna position are also provided for transformation between antenna phase center and center of mass. The real-time ephemerides including broadcast ephemerides and the IGS ultra-rapid products are downloaded via anonymous ftp from ftp://garner.ucsd.edu/. The IGS final products are also downloaded in order to obtain precise reference orbits for accuracy comparison.

An elevation cut-off angle of 5˚ has been applied to reject satellites in poor geometry. The number of tracked GPS satellites above the elevation mask and the position dilution of precision (PDOP) values are shown in Figure 3. The number of available GPS satellites ranges from 6 to 12, and the PDOP value ranges from 1.2 to 3.4. The PDOP value always increases with drop in number of GPS satellites.

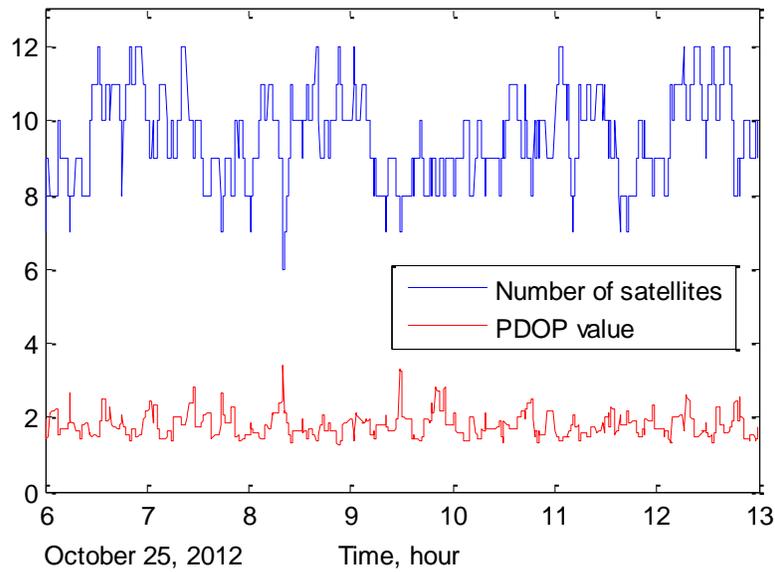

**Fig. 3** Number of tracked GPS satellites above 5° elevation mask and PDOP values

Precise reference orbits and SPP solutions

Before the presentation of kinematic orbit determination results, dynamic orbit determination solutions using a batch least-squares estimator which processes GRAPHIC observations and IGS final products are provided as precise reference for accuracy evaluation. In addition, single point position (SPP) fixes using C/A code and broadcast ephemerides without ionospheric model corrections are also presented for comparison.

The orbit dynamic models employed in the batch estimator are summarized in Table 2. The EGM2008 gravity model truncated at degree and order 100 is used to obtain the accelerations due to the earth's static gravity field. The solid earth tides and ocean tides are modeled to account for tidal effects. The third-body attractions are computed using analytical expansions of solar and lunar coordinates. The maximum area-to-mass ratio of SJ-9A is about 0.016 m$^2$/kg. The accelerations due to atmospheric drag and solar radiation pressure are less than $1 \times 10^{-7}$ m/s$^2$. Thus the non-gravitational accelerations are not modeled. Within this study, piecewise constant empirical accelerations with a 15 min time interval and a constraint of $1 \times 10^{-7}$ m/s$^2$ are used to compensate for unmodeled forces. The GPS measurements are resampled at intervals of 10 s.

Although not depicted here, orbit overlapping analysis indicates a 3D position accuracy of 0.20 m for the batch orbit determination solutions.

Table 2  Summary of dynamic models used in batch orbit determination of SJ-9A

| Item | SJ-9A precise orbit determination |
| --- | --- |
| Gravitational forces | EGM2008 model (100×100) |
| | Solid earth and ocean tides |
| | Solar and lunar gravitational attractions (analytical formulas) |
| Non-gravitational forces | No atmospheric drag and solar radiation pressure model |
| | Piecewise constant empirical acceleration at 15 min interval, constant: $1 \times 10^{-7}$ m/s$^2$ |

The single point positioning algorithm makes use of only C/A code measurements. No empirical models are applied to eliminate the ionospheric effects. The position differences between SPP solutions and the precise reference orbits are shown in Figure 4. The root-mean-square (rms) values of the radial, along-track, cross-track, and 3D position errors are 2.81 m, 0.95 m, 0.87 m, and 3.10 m, respectively. Instant 3-dimensional position errors could reach 9 m at some epochs. The largest part of SPP position errors are attributed to ionospheric delays.

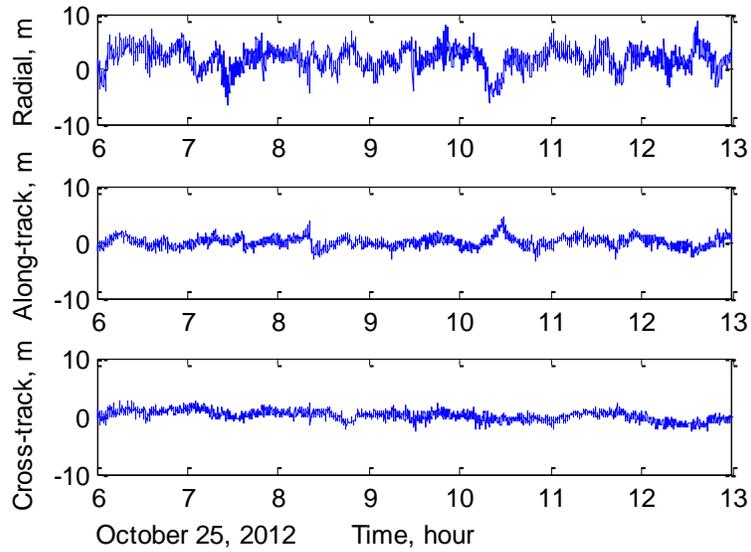

**Fig. 4** Accuracy of SPP positions relative to precise reference orbit in radial, along-track, and cross-track directions. Ionospheric effects are not addressed in SPP

Kinematic orbit determination results

Both the traditional and the improved kinematic orbit determination methods have been tested. The broadcast ephemerides and the IGS ultra-rapid products are used for real-time navigation performance evaluation, whereas the IGS final products are only used for accuracy comparison. The key parameters of the kinematic Kalman filter are set as follows. The sampling rate is 1 Hz. The standard deviation of GRAPHIC observation noise is set to 30 cm. No elevation dependent weighting of GPS data is applied. The standard deviation of the epoch-differenced ionospheric residual is set to 2 cm. The standard deviations of the ambiguity noise when using broadcast ephemerides and IGS ultra-rapid products are both set to 1 mm. A constant ambiguity model is used when using IGS final products. The initial position estimates are obtained from the SPP solutions.

The position accuracies of kinematic orbit determination using IGS final products, IGS ultra-rapid products, and broadcast ephemerides are illustrated in Figures 5-7, respectively. The precise orbit obtained from batch dynamic orbit determination has been used as reference. It is seen that the position solutions obtained from the proposed method are always smoother than those obtained from the traditional method, no matter what kind of ephemeris products are used.

However, the convergence speed can be barely improved. The reason is that the kinematic Kalman filter only constrains the position variations but cannot provide velocity or acceleration information to accelerate the convergence process.

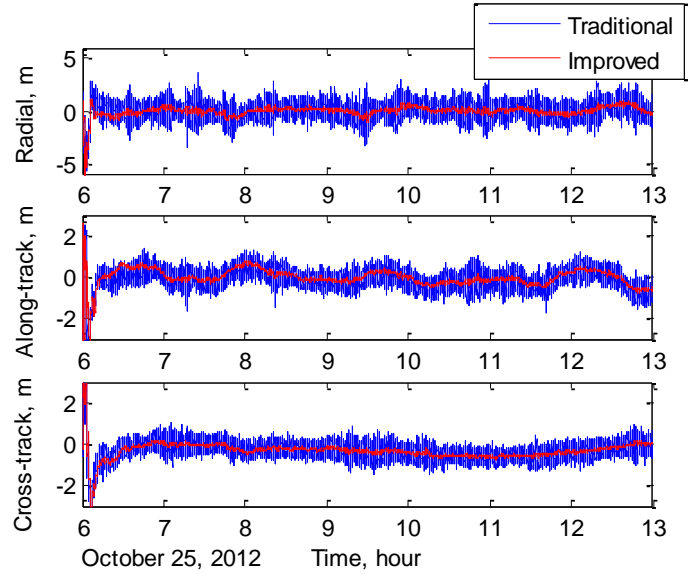

**Fig. 5** Time-varying position errors of traditional and improved kinematic orbit determination using IGS final products

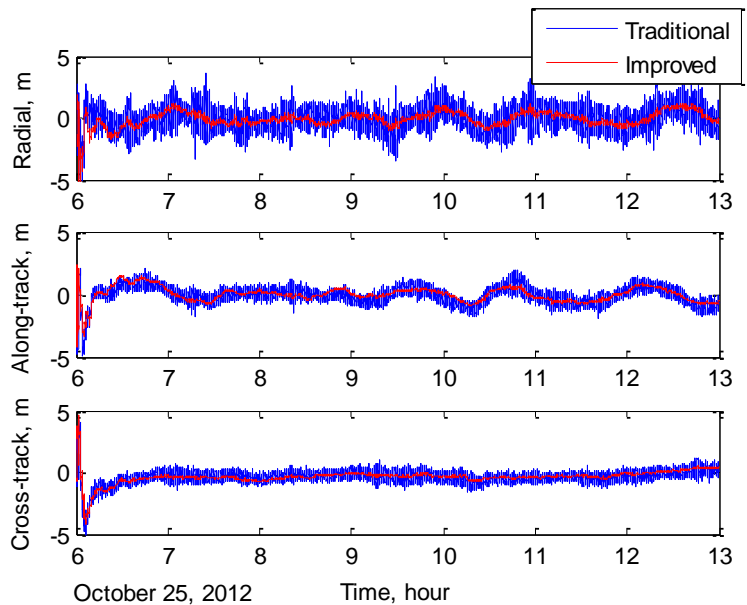

**Fig. 6** Time-varying position errors of traditional and improved kinematic orbit determination using IGS ultra-rapid products

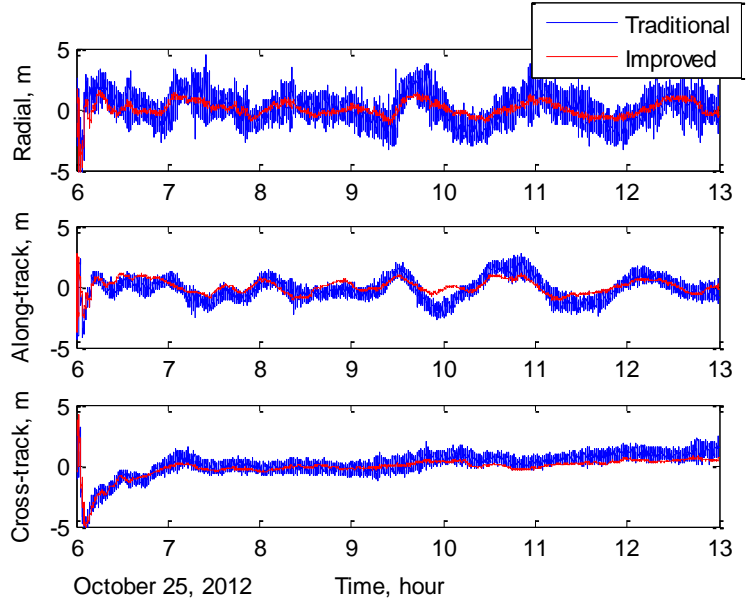

**Fig. 7** Time-varying position errors of traditional and improved kinematic orbit determination using broadcast ephemerides

The statistics of position errors after convergence are summarized in Table 3. First of all, compared with SPP solutions, all the 6 sets of kinematic solutions have better accuracies, since the ionospheric effects are eliminated and the C/A code noise is reduced by a factor of 2 in the GRAPHIC observations. Second, the kinematic Kalman filter effectively reduces the position errors. As stated earlier, the epoch-differenced carrier phase observations are used in the filter to estimate position variations. The estimation errors of position variations are shown in Figure 8. The rms value of estimation errors is 0.010 m. The estimated position variations constrain the absolute position estimates and provide more information for orbit determination. The kinematic Kalman filter behaves like a low-pass filter and significantly suppresses the noise. As reported from Table 3, compared with traditional method, the kinematic Kalman filter reduces the 3D position errors by 38.6%, 31.0%, and 49.0% when using IGS final products, IGS ultra-rapid products, and broadcast ephemerides, respectively. However, the more slowly varying, time-correlated errors are barely suppressed by the proposed method. The reason lies in that systematic measurement errors, such as multi-path effects, phase center variations, and higher-order ionospheric effects, cannot be completely removed in the epoch-differenced observations. The accumulation of residual errors would lead to slowly varying estimation errors. Finally, as expected, the kinematic orbit determination using IGS final products achieves the best accuracy, the IGS ultra-rapid products, and the broadcast ephemerides worst.

Table 3  Statistics of position errors of kinematic orbit determination results

| Ephemerides | | Radial (m) | Along-track (m) | Cross-track (m) | 3D rms (m) | 3D max (m) | Percentage of 3D error > 1 m, % |
|---|---|---|---|---|---|---|---|
| IGS final products | Traditional | 0.033 ±0.664 | -0.083 ±0.399 | -0.303 ±0.330 | 0.899 | 3.837 | 26.3 |
| | Improved | 0.035 ±0.291 | -0.024 ±0.289 | -0.302 ±0.207 | 0.552 | 1.137 | 0.28 |
| IGS ultra-rapid products | Traditional | 0.007 ±0.771 | -0.101 ±0.550 | -0.298 ±0.325 | 1.050 | 3.858 | 40.7 |
| | Improved | 0.022 ±0.458 | -0.046 ±0.425 | -0.274 ±0.237 | 0.724 | 1.355 | 7.06 |
| Broadcast ephemerides | Traditional | 0.053 ±1.077 | -0.195 ±0.900 | 0.358 ±0.538 | 1.559 | 4.841 | 75.3 |
| | Improved | 0.063 ±0.544 | -0.043 ±0.496 | 0.041 ±0.240 | 0.795 | 1.443 | 16.1 |

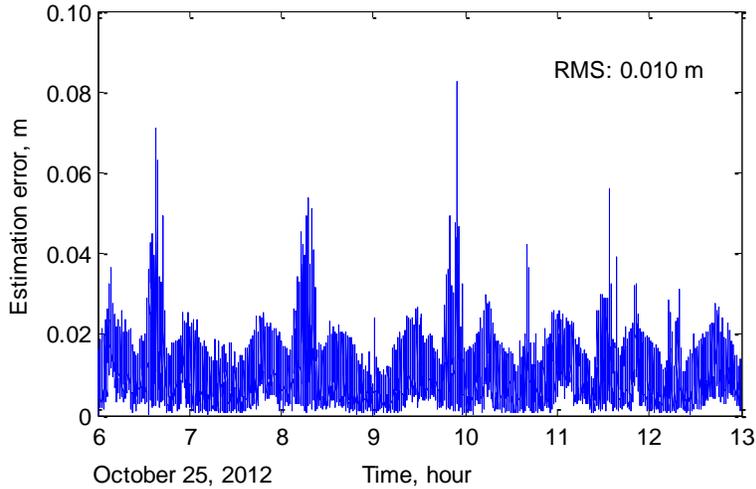

Fig. 8  Estimation errors of position variations

The rms values of 3D position errors obtained from the kinematic Kalman filter are all below 1 m, and the maximum 3D position errors are all below 1.5 m. The percentages of points having a 3D position error above 1 m are 0.28%, 7.06%, and 16.1%. These indicators imply the possibility of the improved kinematic orbit determination method for sub-meter accuracy real-time spacecraft navigation.

In order to investigate the effects of sampling intervals on the performance of the kinematic Kalman filter, three different sampling intervals, i.e., 2 s, 5 s, and 10 s, have also been tested. It is noted that the standard deviation of epoch-differenced ionospheric residual varies with the sampling interval. Within this study, the corresponding standard deviations are set to 4 cm, 10 cm, and 20 cm, respectively. The statistics of position errors are listed in Table 4. With increasing sampling interval, the ionospheric residuals remaining in epoch-differenced observations will get larger. Thus the estimation accuracy of position variations decreases. Accordingly, the absolute position accuracy decreases. As seen from Table 4, the 3D rms errors using a sampling interval of 2 s are slightly larger than those using a 1 s interval. When the sampling interval is increased to 10 s which corresponds to about 60 km separation distance for LEO satellites, the advantage of the proposed method is no longer notable. Therefore, for LEO satellites and vehicles having large velocity, the proposed method is only valid when high sampling rates ($> 0.2$ Hz) are used.

**Table 4** Statistics of position errors with different sampling intervals

| | Sampling interval, s | Radial (m) | Along-track (m) | Cross-track (m) | 3D rms (m) | 3D max (m) | Percentage of 3D error $> 1$ m, % |
|---|---|---|---|---|---|---|---|
| IGS final products | 2 | 0.024 ±0.306 | -0.005 ±0.306 | -0.316 ±0.228 | 0.579 | 1.200 | 7.03 |
| | 5 | -0.009 ±0.429 | -0.047 ±0.401 | -0.308 ±0.347 | 0.750 | 1.905 | 13.5 |
| | 10 | -0.056 ±0.529 | -0.016 ±0.472 | -0.272 ±0.311 | 0.823 | 2.250 | 24.6 |
| IGS ultra-rapid products | 2 | 0.009 ±0.472 | -0.036 ±0.452 | -0.312 ±0.250 | 0.767 | 1.600 | 21.6 |
| | 5 | -0.032 ±0.559 | -0.069 ±0.502 | -0.299 ±0.252 | 0.850 | 2.150 | 28.8 |
| | 10 | -0.110 ±0.565 | -0.058 ±0.468 | -0.382 ±0.438 | 0.855 | 2.127 | 33.6 |
| Broadcast ephemerides | 2 | 0.064 ±0.661 | -0.071 ±0.591 | 0.141 ±0.321 | 0.958 | 1.906 | 44.8 |
| | 5 | 0.026 ±0.860 | -0.138 ±0.776 | 0.313 ±0.258 | 1.235 | 2.894 | 62.1 |
| | 10 | -0.016 ±0.980 | -0.122 ±0.880 | 0.366 ±0.322 | 1.409 | 3.766 | 72.1 |

**Conclusions**

With the improvement of reliability and accuracy, single-frequency GPS receivers have been proposed in recent years for onboard satellite orbit determination. Although the dynamical filtering technique dominates real-time single-frequency navigation, the kinematic method requires lower computational complexity and can be applied for maneuvering spacecraft. In this study, the epoch-differenced carrier phase observations are utilized to improve kinematic orbit determination accuracy.

Based on the experimental results obtained using actual flight data from SJ-9A, it can be concluded that the proposed method effectively reduces position errors. Specifically, large random estimation errors are removed, and most of the position errors are lower than 1 m. With a sampling interval of 1 s, the 3D rms position errors are 0.72 m and 0.79 m for IGS ultra-rapid products and broadcast ephemerides, respectively. These accuracies are comparable to those using dynamic filtering techniques.

The single-frequency GPS receivers have low cost and simple architecture. The use of single-frequency receivers will offer advantages of notable savings in mass, volume, and power for satellite missions and guarantee meter or sub-meter level navigation accuracy at the same time.

**Acknowledgements**

This research was supported by the National Natural Science Foundation of China through cooperative agreement No. 11002008 and has been funded in part by Ministry of Science and Technology of China through cooperative agreement No. 2014CB845303.

**Author Biographies**

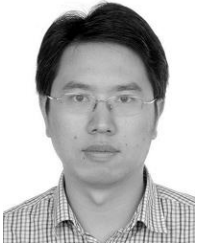

**Pei Chen** received his Ph.D. degree in aerospace engineering from Beihang University, Beijing, in 2008. He is currently an associate professor at the school of astronautics, Beihang University. His current research activities comprise spacecraft navigation, GNSS application, and astrodynamics and simulation.

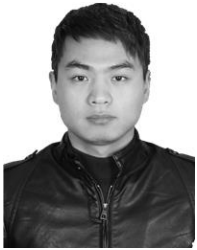

**Jian Zhang** received his B.S. degrees in aerospace engineering from Beihang University, Beijing, in 2015. He is a master student at the school of astronautics, Beihang University. The focus of his current research lies in GNSS navigation.

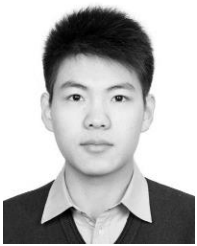

**Xiucong Sun** received his B.S. degree in aerospace engineering from Beihang University, Beijing, in 2010. He is a Ph.D. student at the school of astronautics, Beihang University. The focus of his current research mainly lies in GNSS navigation, spacecraft orbit and attitude determination.